\documentclass[pra,showpacs,showkeys,twocolumn]{revtex4}

\pdfoutput=1

\usepackage{amsmath}
\usepackage{amsfonts}
\usepackage{amssymb}
\usepackage{graphicx}
\usepackage[pdftex]{hyperref}
\hypersetup{
pdfauthor = {Dave Touchette, Haleemur Ali and Michael Hilke},
pdftitle = {5-qubit quantum error correction in a charge qubit quantum computer},
pdfsubject = {Quantum error correction},
pdfkeywords = {Quantum error correction, 5-qubit code, solid-state quantum register, multi charge-qubit system, phase decoherence, relaxation},
pdfcreator = {LaTeX with hyperref package},
pdfproducer = {pdflatex}
}
\numberwithin{equation}{section}
\newcommand{\ket}[1]{\mathop{\left|#1\right>}\nolimits}
\newcommand{\bra}[1]{\mathop{\left<#1\,\right|}\nolimits}
\newcommand{\bk}[2]{\langle #1 | #2 \rangle}
\newcommand{\kb}[2]{| #1\rangle\!\langle #2 |}
\newcommand{\Tr}[2]{\mathop{{\mathrm{Tr}}_{#1}} (#2) }

\def\E{\mathcal{E}}

\begin{document}

\title{5-qubit Quantum error correction in a charge qubit quantum computer}
\author{Dave Touchette, Haleemur Ali and Michael Hilke}
\affiliation{Department of Physics, McGill University, Montreal, Qu\'{e}bec, H3A 2T8, Canada}


\begin{abstract}
    We implement the DiVincenzo-Shor 5 qubit
    quantum error correcting code into a solid-state quantum register. The quantum register is a multi charge-qubit system in a semiconductor environment, where 
    the main sources of noise are phase decoherence and relaxation. We evaluate the decay of the density matrix for this multi-qubit system and perform regular  
    quantum error corrections. The performance of the error correction in this realistic system is found to yield an improvement of the fidelity. The fidelity 
    can be maintained arbitrarily close to one by sufficiently increasing the frequency of
    error correction. This opens the door for arbitrarily long quantum computations.
\end{abstract}
\date{\today}

\maketitle

\section{Introduction}

The task of building a quantum computer, allowing for massive
quantum parallelism \cite{Deu85}, long-lived quantum state superposition and entanglement,
is one of the big challenge of twenty-first century physics. It
requires precise maintenance and manipulation
of quantum systems at a level far from anything that has ever been done before.
Many criteria, nicely summarized by the DiVincenzo criteria \cite{Div00},
need to be satisfied in order to be able to harness the power of
quantum physics to extract powerful information processing,
and while most of these criteria have been satisfied to a reasonable level
by some prospect for the implementation of a quantum information processing
device \cite{NO08}, no such prospect has yet been able to fulfill to a sufficient
level all of these criteria at once. In particular, one criterion that not
many prospects seem to be able to fulfill is that of scalability, the
ability of the model to remain coherent for quantum systems with a large number
of qubits.

Solid-state quantum registers seem to emerge as a medium
that would allow for such scalability, while still being able
to satisfy the other criteria \cite{NO08}. However, as for any physical
implementation of quantum computing, it suffers from noise,
dissipation processes mainly in the form of decoherence and
relaxation \cite{Wei08}. To be able to use the power of quantum computation,
we must be able to eliminate this noise and maintain quantum
information throughout the processing. A general framework for doing so is
quantum error correction (QEC) \cite{Shor95, Stea96}.
It is known that by encoding a single
logical qubit on sufficiently many physical qubits, we can correct
arbitrary errors on a single physical qubit \cite{Shor95}.
But what happens with more realistic multi-qubit errors?

In order to perform a quantum error correction, additional qubits are needed.
However, these additional qubits can lead to a much higher rate of decoherence (even superdecoherence) \cite{superdcoh,IHD05}.
The implementation of the three-qubit repetition code has already
been simulated for a cavity-QED setup \cite{OV10},
showing a significant but limited improvement for the preservation
of quantum information. Here, we consider a realistic noise model
for $N$ charge qubit quantum registers \cite{IHD05},
and simulate this noise model on encoded
qubits to evaluate the efficiency of a QEC
scheme. The quantum error correcting code (QECC) used is
the 5 qubit DiVincenzo-Shor code \cite{DS96}.

We structure this work as follows: we first describe
the representation for quantum systems we will be using,
as well as that for the noise model.
We then further analyze the physical model we study,
and review the QECC we implement within this model.
Section \ref{sec:sim}
then contains the core of this work,
that is, a description of the algorithm used for the
simulation, followed by a discussion of the results obtained.

\section{Representation of quantum systems}
  \label{sec:qusys}

We will be considering the basic unit of quantum information: the qubit.
A qubit can be implemented by any two level quantum system,
and so can be represented by a unit length vector in a two-dimensional
complex Hilbert space, up to an equivalence relation for global phase.
If we consider some prefered orthonormal basis states $\ket{0}$ and $\ket{1}$, a single qubit
can be parametrized by 2 parameters $\theta \in [0, \pi]$ and $\phi \in [0, 2 \pi)$:
\begin{equation}
    \ket{\psi} = \cos\left(\frac{\theta}{2}\right) \ket{0} + e^{i \phi} \sin\left(\frac{\theta}{2}\right) \ket{1}.
\end{equation}
This is the Bloch sphere representation, for pure qubit systems.

When considering multiple interacting qubit systems, the overall system
is described by the tensor product of the component Hilbert spaces.
In general, once two systems have interacted, a complete description
of the whole system cannot be obtained by a description of each subsystem:
this is entanglement. Much of the power of quantum information processing comes from
entanglement, it is a uniquely quantum resource. Note that this is also related
to the difficulty of simulating a multi qubit quantum system:
as a $N$ qubit quantum system cannot be completely described by each of the
$N$ two dimensional subsystems, it must be described by a $2^N$ dimensional system,
so that the memory required to store the information about a $N$ qubit quantum
system scales exponentially with $N$, and so do the time required to simulate
operations on this system.

This also raises the question of the description of subsystems of entangled systems.
It turns out that the density operator formalism \cite{NC00} can solve this issue. In
this formalism, a state $\psi$ is represented by a density matrix
$\rho_\psi =\kb{\psi}{\psi}$ instead of by a state vector $\ket{\psi}$.
Then, if we are only interested in subsequent evolution of a particular
subsystem of the state $\psi$, we can do a partial trace \cite{NC00}
over the other irrelevant subsytems and only evolve the reduced density matrix
of the subsystem of interest. In this way, we obtain the right outcome statistic
on this subsystem while not having to evolve a high-dimensionality state as required
to describe the whole system completely. This density operator formalism also comes
in handy for the description of statistical ensembles $\{ p_i, \ket{\psi_i} \}$, where
the system is prepared in the state $\ket{\psi_i}$ with probability $p_i$.
The corresponding density operator is $\rho = \sum_i p_i \kb{\psi_i}{\psi_i}$,
and here also the evolution of this state leads to the right measurement statistics.

A closely related subject is that of quantum noise. As a quantum system
cannot ever be completely isolated from its environment, they interact and
then become entangled. So, even though the global system-environment pair
undergoes unitary evolution, the main system does not, and this generates quantum
noise. This is well described by the quantum operation formalism \cite{NC00}. Two types of
noise in which we will be mostly interested will be decoherence and relaxation,
and we will see that these are closely related to generalized dephasing channels
 \cite{DS05} and amplitude damping channels respectively \cite{GF05, NC00}.

Since the quantum system of interest will be subject to noise, we want a way to
quantitatively describe how close it will remain to the initial quantum information we
wished to preserve. A good measure of this is the fidelity \cite{Sch96, NC00}, and
for a pure input state $\ket{\psi}$ evolving to a mixed state $\rho$,
the fidelity is given by
\begin{equation}
    F(\ket{\psi}, \rho) = \sqrt{ \bra{\psi} \rho \ket{\psi} }.
\end{equation}

\section{Quantum noise}

The two main types of noise in which we will be interested are
decoherence, which correspond to a loss of quantum coherence
without loss of energy, and relaxation, which correspond to
a loss of energy. We will see that these types of noise are closely
related to well-studied channels: generalized dephasing channels
\cite{DS05, BHTW10} and amplitude damping channels \cite{GF05, NC00}

\subsection{Generalized dephasing channels}

Generalized dephasing channels correspond to physical processes
in which there is loss of quantum coherence without loss of energy.
That is, there exist a preferred basis, called the dephasing basis, such that
pure states in that basis are transmitted without error,
but pure superposition get mixed, and quantum information is
lost to the environment.

If we let $A$ be the input system with orthonormal basis $\{ \ket{i}^A \}$,
$B$ be the output system with orthonormal basis $\{ \ket{i}^B \}$,
and $E$ be the environment with normalized, not necessarily orthogonal
states $\{ \ket{\xi_i}^E \}$, then an isometric map from the input
system to the output-environment system is given by
\begin{equation}
    U^{A \rightarrow BE} = \sum_i \ket{i}^B \ket{\xi_i}^E \bra{i}^A.
\end{equation}
Then, starting with an input state $\rho^A$, we get
the output state $\sigma^B$ by first applying the
isometry $U^{A \rightarrow BE}$, then tracing over $E$:
\begin{equation}
    \sigma^B = \Tr{E}{U \rho^A U^\dagger}
             = \sum_{i, j} \bra{i} \rho^A \ket{j} \kb{i}{j}^B \bk{\xi_j}{\xi_i}.
\end{equation}
As we can see, in this representation, the output corresponds to the Hadamard
product between the $(\bra{i} \rho^A \ket{j})$ input matrix and
the $(\bk{\xi_i}{\xi_j}^{\dagger})$ decoherence matrix. Also, since the ${ \ket{\xi_i} }$
are normalized, $\bk{\xi_i}{\xi_i} = 1$, the diagonal terms
are left unchanged:
\begin{equation}
    \bra{i} \sigma^B \ket{i} = \bra{i} \rho^A \ket{i}.
\end{equation}
These diagonal terms correspond
to the probability of measuring the quantum state in the corresponding
dephasing basis state, while the off-diagonal terms are coherence (phase) terms.
For these off-diagonal terms, the Cauchy-Schwarz inequality
($|\bk{\xi_i}{\xi_j}|^2 \leq \bk{\xi_i}{\xi_i} \bk{\xi_j}{\xi_j} = 1$)
tells us that the output terms must be smaller than or
equal to the input terms: $|\bra{i} \sigma^B \ket{j}| \leq |\bra{i} \rho^A \ket{j}|$.
In the special case where the $\{ \ket{\xi_i}^E \}$ are also orthogonal, we
obtain a completely dephasing channel, which kills all off-diagonal terms,
$\bra{i} \sigma^B \ket{j} = 0$ if $i \neq j$.
For a subsequent measurement in the dephasing basis, this yields
a classical statistical output with no quantum coherence terms:
$\sigma^B = \sum_i \bra{i} \rho^A \ket{i} \kb{i}{i}^B$.

\subsection{Amplitude damping channels}

Amplitude damping channels correspond to physical processes in which
there is a loss of energy from the quantum system of interest to
the environment. That is, there is a probability that a quantum
state passes from a higher energy excited state to a lower energy state.
For the two-dimensional quantum systems in which we are mostly interested,
this corresponds to a probability of passing from the excited state to the
ground state, while the ground state is left unaffected. In the quantum
operation formalism, this channel has operation elements $\{ E_0, E_1 \}$:
\begin{equation}
  \label{eq:adc}
    E_0 = \left(
    \begin{array}{cc}
        1 & 0  \\
        0 & \sqrt{1-\gamma}
    \end{array}
    \right),
    E_1 = \left(
    \begin{array}{cc}
        0 & \sqrt{\gamma}  \\
        0 & 0
    \end{array}
    \right),
\end{equation}
where the matrix representation is given in the energy eigenbasis,
and the parameter $\gamma$ corresponds to the probability of decay.
Then, for some input state $\rho^A$, we get an output state $\sigma^B$:
\begin{equation}
    \sigma^B = \sum_{i = 0, 1} E_i \rho^A E_i^\dagger.
\end{equation}
If we represent the action of the channel as $\E$, i.e.
$\E (\rho^A) = \sigma^B$, we can extend this definition
to multi-qubit systems. Making the assumption that energy lost for
each two-dimensional subsystem is independent, a tensor product of
this channel, $\E^{\otimes n}$, corresponds to a loss of energy in
each subsystem independently.

\section{Decoherence and relaxation in charge qubit registers}

\begin{figure}[h!]
  \centering
    \includegraphics[width=0.5\textwidth]{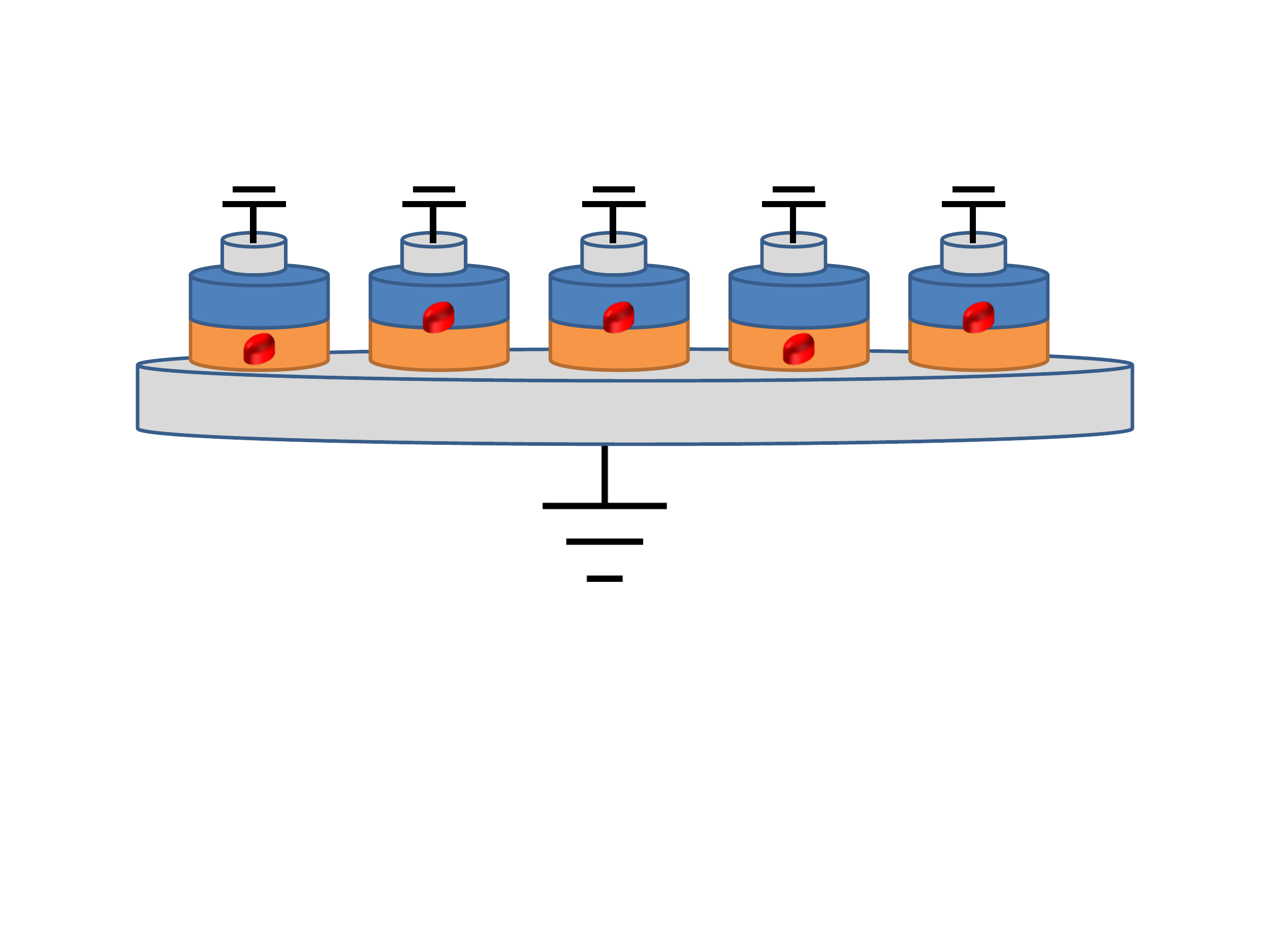}
         \vspace{-3cm}
   \caption{Physical model for the 5-qubit quantum register. The qubit geometry is inspired from recent experiments on coherent level mixing in vertical quantum 
		dots \cite{Chris}.}
  \label{fig:qubit}
\end{figure}

Solid state quantum systems represent an interesting prospect
for the physical implementation of scalable quantum devices \cite{NO08}.
A lot of research goes into implementation of superconducting
qubits \cite{MSS01, WS05}. Another important avenue of research
are quantum dots, in which two level quantum systems can
be implemented in either the spin or position degree
of freedom of the electron trapped in the dots. While spin
quantum dots have a longer decoherence time \cite{Elz04, Pet06}, spins are also more difficult to manipulate. In charge quantum dots,
where the basis states correspond to the position of the electron in either of two adjacent
quantum dots, the coupling is strong and hence fast manipulation is possible, but this also leads to shorter decoherence times.
We will consider such a system, which offers an interesting playground for current technology research \cite{Sh07, Hay03}.
Moreover, the decoherence for $N$ qubit quantum registers,
which can be seen in FIG. \ref{fig:qubit},
has already been computed by Ischii et al. \cite{IHD05}, and
it is this model which is used in this simulation.

\begin{figure}[h!]
  \centering
    \includegraphics[width=0.5\textwidth]{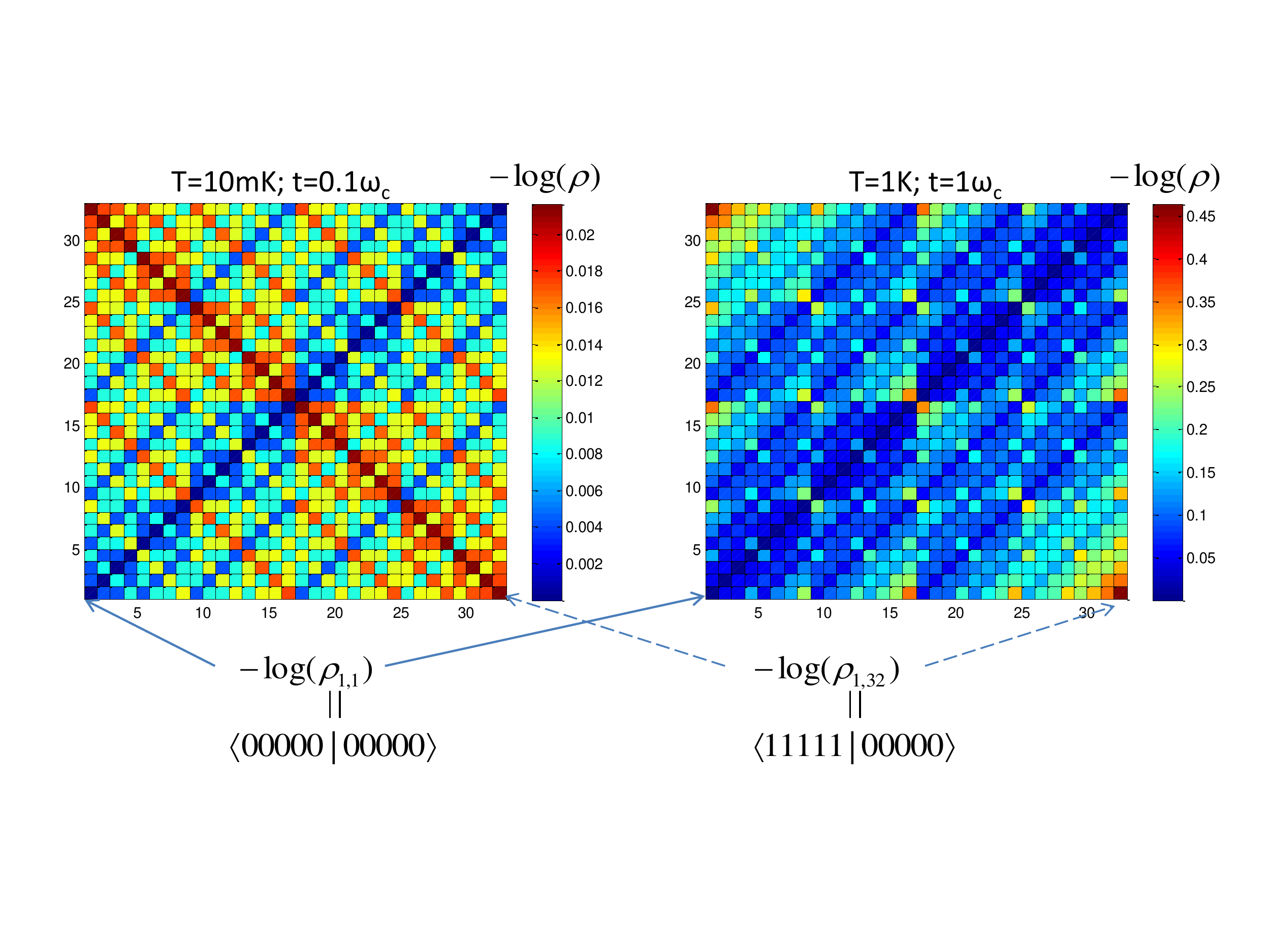}
     \vspace{-1.5cm}
  \caption{(Color online) Matrix elements of the decay function density matrix due to decoherence.}
  \label{fig:rho}
\end{figure}

It is found that for these charge quantum dots, the decoherence
of the quantum system can be represented as a generalized dephasing
channel in the canonical basis of the charge qubits, corresponding
to the position eigenstates. Indeed, pure states in that basis are
perfectly transmitted, while pure superpositions become noisy.
It is then possible to compute the evolution of the quantum
register by taking the Hadamard product of the input
density matrix with a decoherence matrix computed in \cite{IHD05}, thus getting
the decohered output state. We chose as physical timescale unit $\omega_0$, which corresponds to the upper cut-off frequency.
In charge qubits in GaAs we typically have
\begin{equation}
    \omega_0 = 0.2 \times 10^{-10} s.
\end{equation}
This leads to the decay of the off-diagonal elements of the N-qubit density matrix.
The structure of this decay is non-trivial and is shown in FIG. \ref{fig:rho}.
In general, this leads to an error, which cannot be reduced to a single qubit error.

In addition, this decoherence model does not take into account any
relaxation the system might be subject to. We assume that
the relaxation is independent of decoherence, and
also that energy loss for each qubit is independent.
For each two dimensional quantum subsystem, if $\ket{0}$
and $\ket{1}$ correspond to the two position eigenstates, then
the two energy eigenstates are $\ket{+} = \frac{1}{\sqrt{2}} (\ket{0} + \ket{1})$
for the ground state, and $\ket{-} = \frac{1}{\sqrt{2}} (\ket{0} - \ket{1})$
for the excited state. We can then compute the effect of relaxation by
operating a tensor product of amplitude damping channels to each
qubit in the register, where the operation elements
(\ref{eq:adc}) are in the $\{ \ket{+}, \ket{-} \}$ basis.
The decay parameter $\gamma$ is time $(t)$ and temperature
$(T)$ dependent,
\begin{equation}
    \gamma = 1 - exp(- t \cdot T),
\end{equation}
setting $k_B = \hbar =1$.

\section{Quantum error correction with a perfect 5 qubit code}

Since our quantum register is subject to noise, we want to encode an input
quantum state in such a way that we can preserve the quantum information
even when the physical system undergoes noise. A general framework for doing
this is quantum error correction (QEC) \cite{Shor95, Stea96},
which encode a logical qubit containing
the quantum information on multiple physical qubits, and then
usually by performing syndrome measurements on these physical qubits, we
can determine a certain set of errors, and apply the corresponding correction \cite{NC00}.
Many quantum error correcting codes (QECC)
have been found that can correct an arbitrary error on a single
physical qubit, and it is known that the smallest such codes require
5 physical qubits \cite{LMPZ96}.

One of these 5 qubit codes is the DiVincenzo-Shor QECC \cite{DS96}.
The code words for the logical states $\{ \ket{0}_L, \ket{1}_L \}$ are \cite{NO08}:
\begin{eqnarray}
  \label{eq:code}
    \ket{0}_L &=& \frac{1}{4} \big( \ket{00000} + \ket{11000} + \ket{01100} + \ket{00110} \nonumber \\
              &+& \ket{00011} + \ket{10001} - \ket{10100} - \ket{01010} \nonumber \\
              &-& \ket{00101} - \ket{10010} - \ket{01001} - \ket{11110} \nonumber \\
              &-& \ket{01111} - \ket{10111} - \ket{11011} - \ket{11101} \big) \nonumber \\
    \ket{1}_L &=& \frac{1}{4} \big( \ket{11111} + \ket{00111} + \ket{10011} + \ket{11001} \nonumber \\
              &+& \ket{11100} + \ket{01110} - \ket{01011} - \ket{10101} \nonumber \\
              &-& \ket{11010} - \ket{01101} - \ket{10110} - \ket{00001} \nonumber \\
              &-& \ket{10000} - \ket{01000} - \ket{00100} - \ket{00010} \big) \nonumber
\end{eqnarray}
with stabilizers \cite{Gott96}:
\begin{eqnarray}
    M_0 = I \otimes Z \otimes X \otimes X \otimes Z \nonumber \\
    M_1 = Z \otimes I \otimes Z \otimes X \otimes X \nonumber \\
    M_2 = X \otimes Z \otimes I \otimes Z \otimes X \\
    M_3 = X \otimes X \otimes Z \otimes I \otimes Z, \nonumber
\end{eqnarray}
which leave the code invariant.
The $X$ and $Z$ operators are respectively the $X$ and $Z$ Pauli operators.
The encoding circuit can be seen in \cite{NO08}. This is
a perfect non-degenerate code, meaning that all single qubit errors map
to a different syndrome, as can be seen in TABLE \ref{tb:error}.
To obtain a measurement of the error syndrome, we measure observables in the
eigenbasis of each of the four stabilizers in (\ref{eq:code}), and
depending on the outcome of these four measurements
we apply the corresponding correction from TABLE \ref{tb:error}.

An alternative to these multi-qubit measurements \cite{NC00} is to use ancillary qubits to
conditionnally apply each of the four stabilizers on a $\ket{+}$ state, thus
recording the phase (M0-M3 have eigenvalues $\pm 1$). Measuring
these four ancilla qubits in the $\{ \ket{+}, \ket{-} \}$ basis then gives the same result
as above without having to perform multi-qubit measurements. This syndrome
measurement circuit is given in \cite{NO08}.

\begin{table}[h!]
  \begin{tabular}{|c|c|c|c|c|c|c|c|c|c|c|c|c|c|c|c|c|}
   \hline
     $M$  &  0  &   1   &   2   &   3   &   4   &   5   &   6   &   7   &   8   &   9   &  10   &  11   &  12   &  13   &  14   &  15   \\
    $E_M$ & $I$ & $Z_2$ & $X_0$ & $Z_3$ & $X_3$ & $X_1$ & $Z_4$ & $Y_3$ & $Z_1$ & $X_4$ & $X_2$ & $Y_2$ & $Z_0$ & $Y_1$ & $Y_0$ & $Y_4$ \\
   \hline
  \end{tabular}
  \caption{Table of the different possible single qubit errors $E_M$,
                with their associated syndrome measurement output $M$, which has
                binary expension $M_3 M_2 M_1 M_0$.}
  \label{tb:error}
\end{table}

This code is guaranteed to correct any single qubit error, but
it is not designed to correct multi-qubit errors, which our noise model
will introduce. It will then be interesting to see how well this code
will perform for an approximate correction of multi-qubit errors.

\section{Quantum error correction in a charge qubit quantum computer}
  \label{sec:sim}

The qubit model for physical implementation we consider in this simulation
is a charge quantum dot. The two principal sources of noise with this model are
decoherence and relaxation.

The effect of decoherence on a $N$-qubit register had been computed in \cite{IHD05},
where decoherence is caracteristic of a generalized dephasing channel \cite{DS05},
leaving computational basis states unaffected.

The effect  of relaxation correspond to a probabilistic loss of energy,
where the excited state correspond to the $\ket{-}$ state,
and the ground state correspond to the $\ket{+}$ state.
We make the assumption that this relaxation for each qubit is independant,
so that we implement this noise has a tensor product of an amplitude damping channel
in the $\{ \ket{+}, \ket{-} \}$ basis, on each qubit.

To counter the effect of noise, we encode a logical qubit into 5 qubits,
using the DiVincenzo-Shor QECC, which can perfectly correct errors on a single qubit.
However, the noise model used creates errors on multiple qubits, and so it is not obvious
that QEC with the DiVincenzo-Shor code can extend the lifetime
of our logical qubit.

Here, we verify the effect of such a quantum error correction scheme,
for different input states and different frequencies of correction.
The algorithm used for the simulation is presented first, and the
results of the simulation are presented next.

\subsection{Implementation of QEC algorithm}

\begin{figure}[h!]
  \centering
    \includegraphics[width=0.5\textwidth]{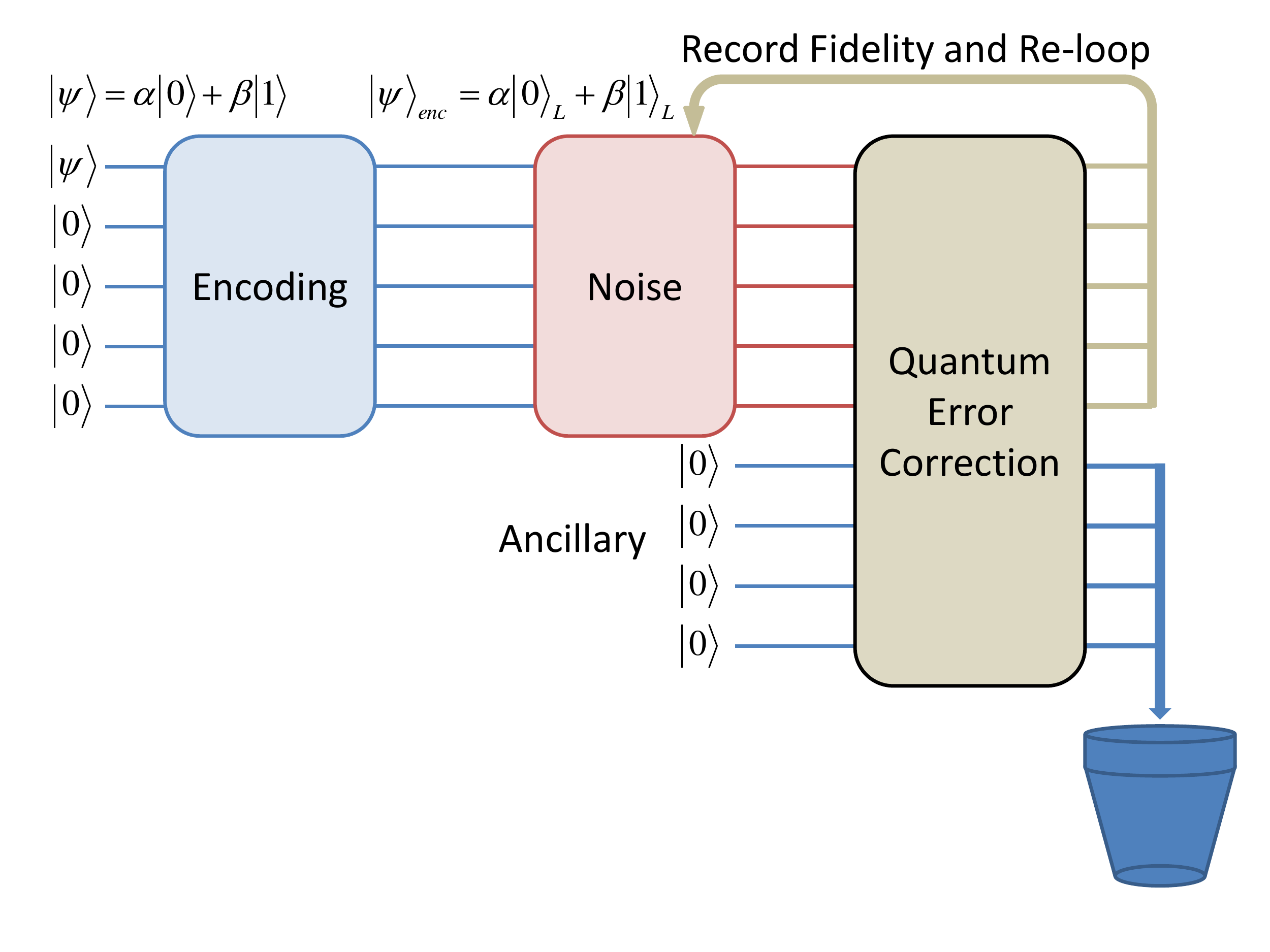}
  \caption{Overview of the algorithm for the simulation of quantum error correction
                      of a solid-state quantum register.}
  \label{fig:algo}
\end{figure}

An overview of the algorithm can be seen in FIG. \ref{fig:algo}.
First, the logical qubit is encoded into 5 physical qubits using
the DiVincenzo-Shor QECC. This encoding is assumed to be perfect.
We only consider pure input qubits, which are completely caracterized by two
real parameters corresponding to their position on the Bloch sphere:
$\theta \in [0, \pi]$, $\phi \in [0, 2 \pi)$.

After encoding, the quantum register is subjected to noise for a given period of
time $t$, depending on the frequency of error correction. In our simulation, the decoherence noise and the relaxation noise are applied successively, and we
checked numerically that the order does not affect the result. The intensity of the noise depends
on the temperature $T$ of simulation, and the probability of relaxation is adjusted such
that without error correction, after a time $\omega_0$ the fidelity of the noisy state with respect to the
pure input state is the same as that of decoherence.

Quantum error correction is then performed. Here we assume perfect operation of each component
of the QEC algorithm.
However, instead of performing the measurement of the syndrome and then applying
the corresponding correction (see TABLE \ref{tb:error}), we defer the measurement to the end
of the circuit and perform conditional quantum correction \cite{NC00, NO08}.
To do so, four ancillary qubits are used to record the syndrome measurements.
It is possible to avoid multi-qubit measurements by using these ancillary
qubits, but the drawback is that we pass from a 5 qubit quantum system to a
9 qubit quantum system. This increases the difficulty of physical implementation
because of the additionnal qubits, and also because every round of QEC
requires four fresh ancilla qubits. This also slows the simulation a lot, since
we pass from a $2^5 = 32$ dimensional complex vector space to a
$2^9 = 512$ one, and the matrix operations are accordingly scaled.

After performing the conditional QEC, these ancilla qubits are traced over,
which corresponds to a measurement without a record of the outcome. This gives an
output state which is a statistical ensemble corresponding to the different possible corrected
states.

To perform the conditional
quantum error correction, we implement a quantum circuit
which would correct for the different errors in TABLE \ref{tb:error},
corresponding to the value of the ancilla qubits. If we let $\ket{M}$ correspond
to the ancilla state with binary expansion $M_3 M_2 M_1 M_0$, for $M$ running
from $0$ to $15$, and if $E_M$ is the corresponding error, then the QEC circuit is
\begin{equation}
    \prod_{M = 0}^{15} \big( E_M^\dagger \otimes \kb{M}{M} + I_5 \otimes ( I_4 - \kb{M}{M} ) \big),
\end{equation}
where the order in the operator product is irrelevant since the terms commute.

Following the quantum error correction step, a record of the fidelity of this output state compared
to the input state is taken. We then reloop over the noise, quantum error correction and
fidelity recording steps, until the sum of all the noise time steps add up to the desired total
time of the simulation.

\subsection{QEC performance}

\subsubsection{Decoherence}

\begin{figure}[h!]
  \centering
    \includegraphics[width=0.4\textwidth]{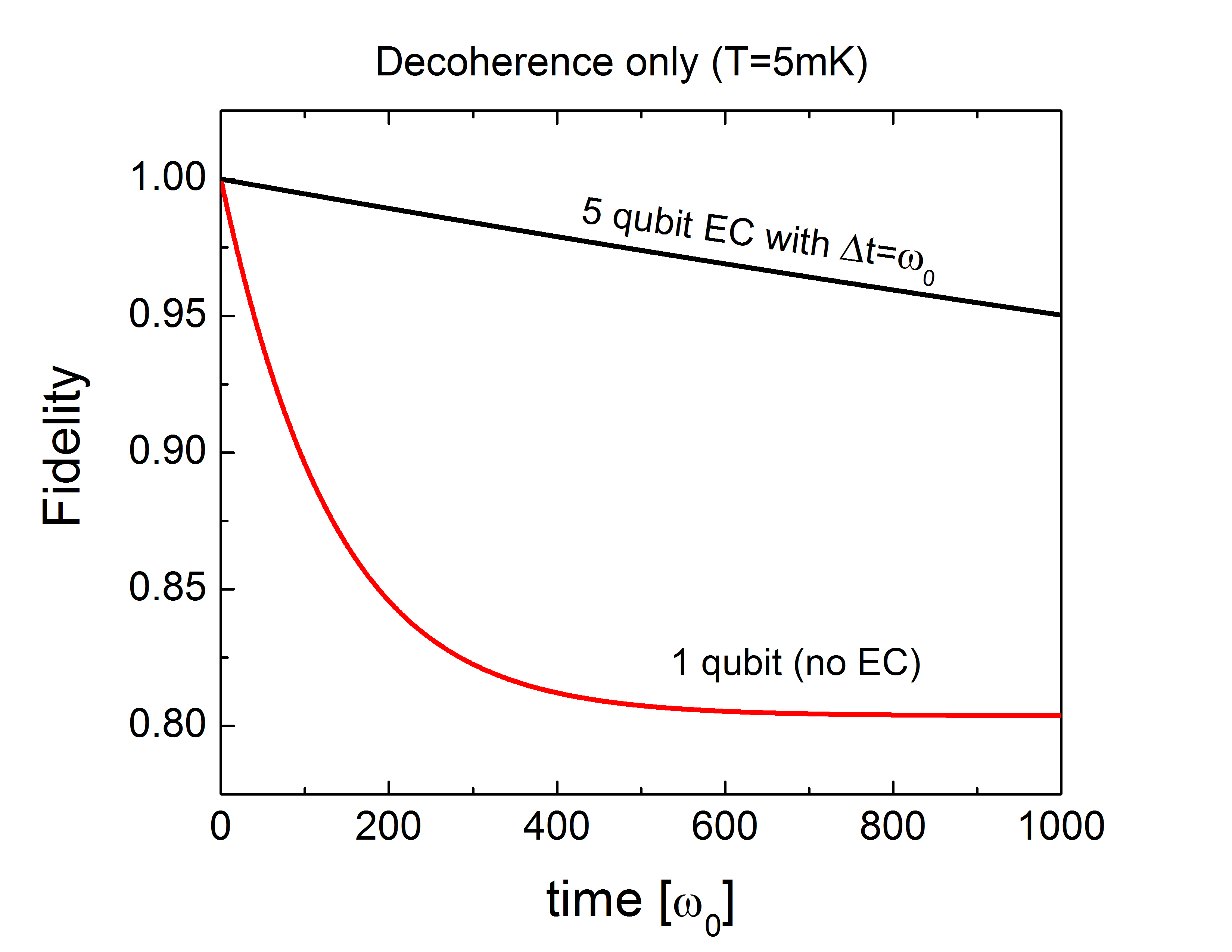}
  \caption{Simulation of quantum error correction with only decoherence. The time between two successive QECs is $\Delta t=\omega_0$.
                    For a total of $1000 \omega_0$, the top curve shows the result of the QEC and the non-corrected 1qubit decoherence is shown in the bottom curve.
                    The input qubit is
            $\ket{\psi} = \cos\left(\frac{1}{2}\right) \ket{0} + e^{i} \sin\left(\frac{1}{2}\right) \ket{1}$
            and the temperature is assumed to be 5 mK.}
  \label{fig:decoh}
\end{figure}

We can see on FIG. \ref{fig:decoh} the results of the simulation
when our quantum register is subject only to decoherence.
With a small frequency of error correction of $\omega_0^{-1}$,
the fidelity is already improved over the uncorrected qubit.
For a basis qubit, either $\ket{0}$ or $\ket{1}$,
we do not have decoherence, but the encoded state does suffer
from decoherence. The encoded basis states however do exhibit
an interesting behavior when they are decohered, but not error corrected.
In fact, even though the noise takes the 5 qubit state out of the code,
when decoded we still get back the original basis state. For these
basis states, we could understand this behavior by noting that the code
words for the logical $\ket{0}$ and $\ket{1}$ contain different 5 qubit
basis states, as we can see in (\ref{eq:code}). But for pure
superposition of these basis states, if we do not error correct
the encoded states, it also decoheres in a way such that once decoded,
the same state as the unencoded decohered state is obtained, which is even more surprising.

\subsubsection{Relaxation}

\begin{figure}[h!]
  \centering
    \begin{tabular}{cc}
    \includegraphics[width=0.25\textwidth]{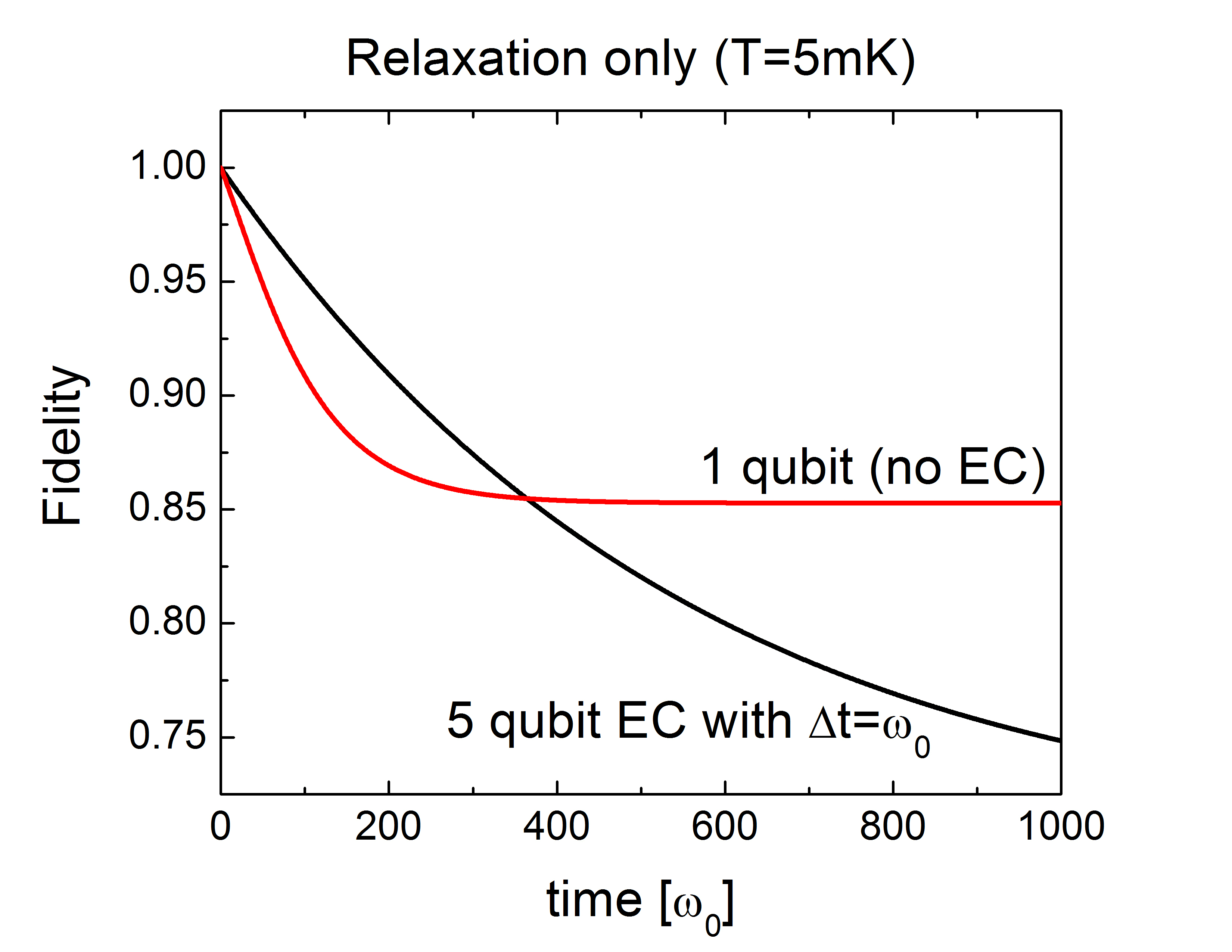} & \includegraphics[width=0.25\textwidth]{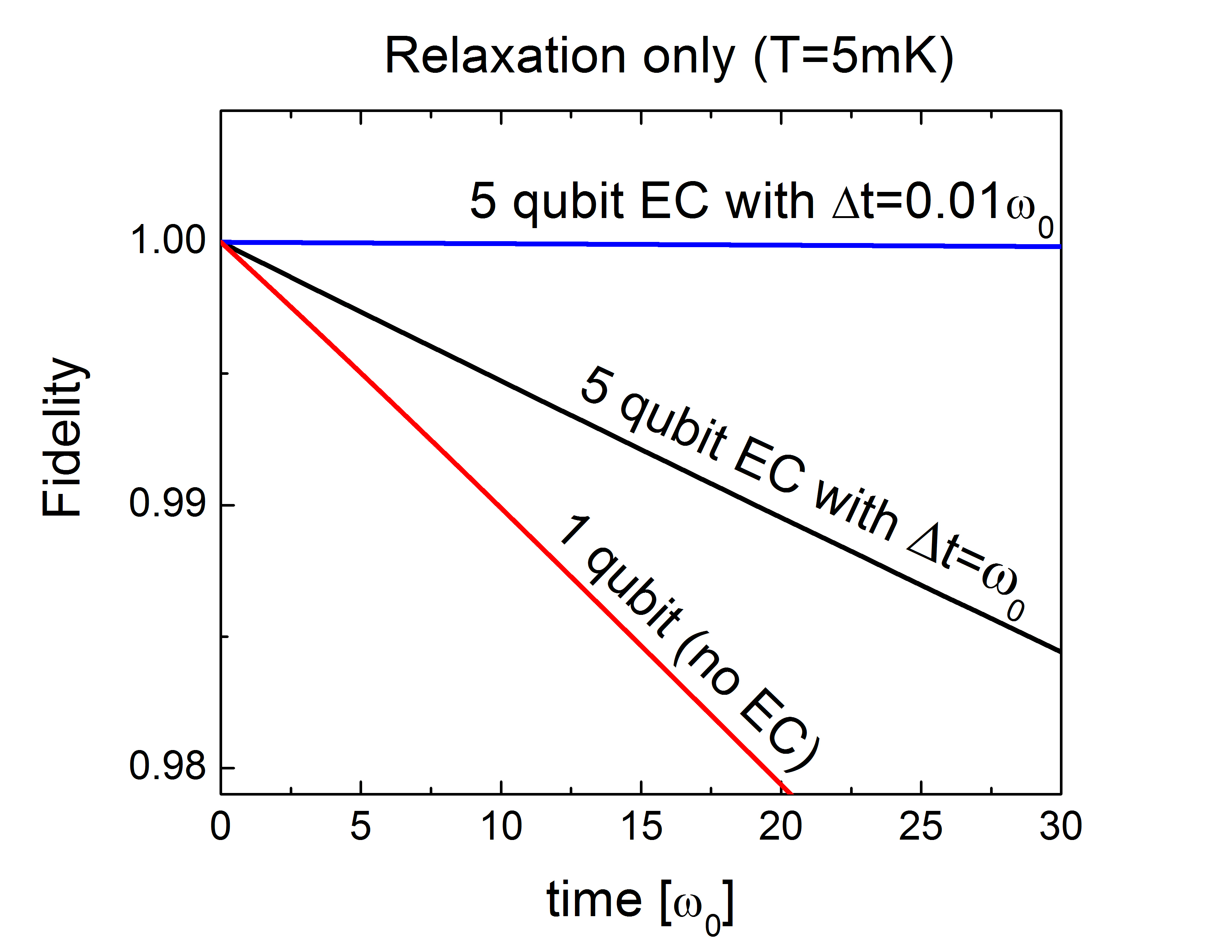}
    \end{tabular}
  \caption{Simulation of quantum error correction with only relaxation: In the left graph the fidelity as a function of time is shown for the same $\Delta 
            t=\omega_0$ as in FIG. \ref{fig:decoh}. By reducing the time interval between two QECs the fidelity remains close to one as seen in the right graph.
                    For both graphs, the input qubit is the same as in FIG. \ref{fig:decoh}.}
  \label{fig:relax}
\end{figure}

We can see in FIG. \ref{fig:relax} the results of the simulation
when we only consider relaxation affecting our quantum register. This
type of noise has a stronger temperature dependency and is
stronger than decoherence for temperatures above 5 mK, at which point we
obtain on average a similar drop in fidelity at 1 $\omega_0$. It is
also harder to correct than decoherence. We thus need a much higher
frequency of quantum error correction to keep the fidelity close
to 1 until we have reached a saturation in relaxation, that is we
have reached the lowest energy state for the unencoded qubit.
This low energy state is the only state invariant under action of the channel.

\subsubsection{Simultaneous decoherence and relaxation}

\begin{figure}[h!]
  \centering
    \includegraphics[width=0.4\textwidth]{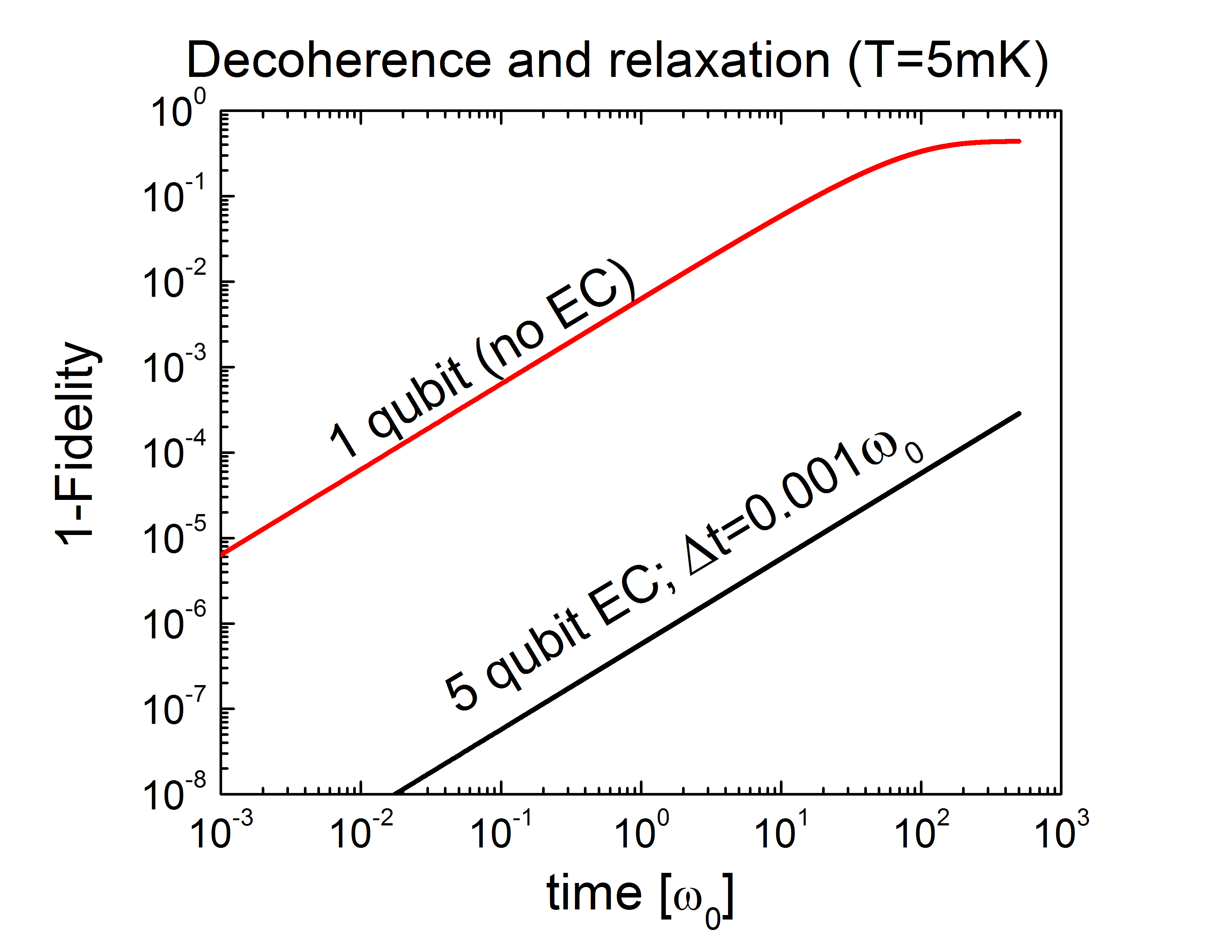}
  \caption{1 minus fidelity is shown as a function of time for a very short $\Delta t=0.001 \omega_0$.
                The high-frequency QEC for joint decoherence and relaxation enables the quantum information
                to be preserved almost perfectly.
                The input qubit is the same as in FIG. \ref{fig:decoh}.}
  \label{fig:scale}
\end{figure}

When combining the effect of decoherence and relaxation, the frequency
of error correction is on the order of what is needed for relaxation,
since relaxation is much harder to correct than decoherence. Here too, when taking into account decoherence and relaxation, we are
still able to maintain the fidelity arbitrarily close to one (see FIG. \ref{fig:scale}). This tells us that even though it may require a high frequency of quantum
error correction, the 5 qubit code is able to correct a realistic
multi-qubit error system.

\subsection{Discussion}

We have only considered noise in the N-qubit system.
That is, an important assumption made for our simulation is that of a perfect operation of the QEC algorithm.
However, in a realistic fault-tolerant quantum computation (FTQC) \cite{Shor96} ,
errors will also occur within the operation of the gates and measurements,
and this also during the QEC steps. Similar concerns led some
physicists to wonder if the error model adopted to get to the
threshold theorem \cite{KL96, AB96} for FTQC
is physical enough so that FTQC is possible at all \cite{Dya06}.
Hence, such a simulation, with a realistic model for operations, would be an important
step toward determining if physical implementations of quantum computers are possible.
It is possible that these types of errors can be corrected in a similar manner, but this is beyond the scope of this work.

The model considered introduces multi-qubit errors, which are much smaller than full single-qubit ones when the time evolution is very short.
At the lowest order, we showed that these small multi-qubit errors can be corrected by single-qubit QEC codes.
However, this increases the required QEC frequency,
but it is certainly possible to find other error correcting schemes optimized for multi-qubit errors, which would require a lower QEC frequency. 

Also, since the decoherence time for charge qubits is short,
the frequency of QEC required to maintain good fidelity for extended periods
is too high to be feasible experimentally with current techniques. However, this is a proof of principle and it is reasonable to expect that
for other implementations of qubits like spin qubits
or superconducting qubits, where decoherence times can be much longer, similar performances would require a
QEC frequency that is attainable experimentally.

\section{Conclusion}

We have simulated the implementation of the 5 qubit DiVincenzo-Shor QECC
for a charge qubit quantum register. The register was subjected to a realistic
noise model, consisting of decoherence and relaxation.
Even though the QECC is designed to perfectly correct only single qubit errors,
an application of the QEC routine with a high enough frequency was able
to limit the effect of multi-qubit errors introduced by our noise model.
In fact, by adjusting the frequency of error correction, we can get
the fidelity as close to one as possible, and for extended periods of time.
When considering both types of noise separately, relaxation showed to be
harder to error correct than decoherence, requiring higher frequency of
QEC for the same results.

\section*{Acknowledgement}
M.H. acknowledges financial support from INTRIQ.
D.T. acknowledges financial support from a NSERC USRA for the duration of
this work.



\begin{thebibliography}{99}

\bibitem {Deu85} D. Deutsch,
\textit{Quantum theory, the Church-Turing principle and the universal quantum computer},
Proc. R. Soc. Lond. A, 400:97 (1985).

\bibitem {Div00} D. P. DiVincenzo,
\textit{The Physical Implementation of Quantum Computation},
Fortschr. Phys. 48, 771 (2000).
arXiv:quant-ph/0002077v3

\bibitem {NO08} M. Nakahara and T. Ohmi,
\textit{Quantum Computing: From Linear Algebra to Physical Realizations},
(CRC Press, 2008).

\bibitem {Wei08} U. Weiss,
\textit{Quantum dissipative systems},
(World Scientific publishing co., third ed., 2008).

\bibitem {Shor95} P. W. Shor,
\textit{Scheme for reducing decoherence in quantum computer memory},
Phys. Rev. A 52, 2493 (1995).

\bibitem{Stea96} A. Steane,
\textit{Simple Quantum Error Correcting Codes},
Phys. Rev. A 54, 4741 (1996).
arXiv:quant-ph/9605021v1

\bibitem{superdcoh} G. M. Palma, K. A. Suominen and A. K. Ekert,
\textit{Quantum computers and dissipation},
Proc. Roy. Soc. Lond. A 452, 567 (1996).
arXiv:quant-ph/9702001v1

\bibitem{IHD05} B. Ischi, M. Hilke and M. Dube,
\textit{Decoherence in a N-qubit solid-state quantum register},
Phys. Rev. B 71, 195325 (2005).
arXiv:quant-ph/0411086v2

\bibitem{OV10} C. Ottaviani amd D. Vitali,
\textit{Implementation of a three-qubit quantum error-correction code in a cavity-QED setup},
Phys. Rev. A 82, 012319 (2010).
arXiv:quant-ph/1005.3072v2

\bibitem {DS96} D. P. DiVincenzo and P. W. Shor,
\textit{Fault-Tolerant Error Correction with Efficient Quantum Codes},
Phys. Rev. Lett. 77, 3260 (1996).
arXiv:quant-ph/9605031v2

\bibitem {NC00} M. A. Nielsen and I. L. Chuang,
\textit{Quantum Computation and Quantum Information},
(Cambridge Univ. Press, 2000).

\bibitem {DS05} I. Devetak and P. W. Shor,
\textit{The capacity of a quantum channel for simultaneous transmission of classical and quantum information},
Commun. Math. Phys. 256, 287 (2005) [ISI].
arXiv:quant-ph/0311131

\bibitem{GF05} V. Giovannetti and R. Fazio,
\textit{Information-capacity description of spin-chain correlations},
Phys. Rev. A 71, 032314 (2005).
arXiv:quant-ph/0405110v3

\bibitem{Sch96} B. W. Schumacher,
\textit{Sending entanglement through noisy quantum channels},
Phys. Rev. A 54, 2614 (1996).
arXiv:quant-ph/9604023v1

\bibitem{BHTW10} Kamil Bradler, Patrick Hayden, Dave Touchette, and Mark M. Wilde,
\textit{Trade-off capacities of the quantum Hadamard channels},
Phys. Rev. A 81, 062312 (2010).
arXiv:quant-ph/1001.1732v2

\bibitem{MSS01} Y. Makhlin, G. Schon and A. Shnirman,
\textit{Quantum-state engineering with Josephson-junction devices},
Rev. Mod. Phys. 73, 357 (2001).
arXiv:cond-mat/0011269v1

\bibitem{WS05} G. Wendin and V. S. Shumeiko,
\textit{Superconducting Quantum Circuits, Qubits and Computing},
Prepared for Handbook of Theoretical and Computational Nanotechnology (2005).
arXiv:cond-mat/0508729v1

\bibitem{Elz04} J. M. Elzerman, R. Hanson, L. H. Willems van Beveren, B. Witkamp, L. M. K. Vandersypen and L. P. Kouwenhoven,
\textit{Single-shot read-out of an individual electron spin in a quantum dot},
Nature 430, 431 (2004).
arXiv:cond-mat/0411232v2

\bibitem{Pet06} J.R. Petta, A.C. Johnson, J.M. Taylor, E.A. Laird, A. Yacoby, M.D. Lukin, C.M. Marcus, M.P. Hanson and A.C. Gossar,
\textit{Preparing, manipulating, and measuring quantum states on a chip},
Physica E 35, 251 (2006).

\bibitem{Sh07} G. Shinkai, T. Hayashi, Y. Hirayama and T. Fujisawa,
\textit{Controlled resonant tunneling in a coupled double-quantum-dot system},
Appl. Phys. Lett. 90 103116 (2007).

\bibitem{Hay03} T. Hayashi, T. Fujisawa, H. D. Cheong, Y. H. Jeong, and Y. Hirayama  
\textit{Coherent Manipulation of Electronic States in a Double Quantum Dot},
Phys. Rev. Lett. 91, 226804 (2003).
arXiv:cond-mat/0308362v1

\bibitem{Chris} C. Payette, S. Amaha, G. Yu, J. A. Gupta, D. G. Austing, S. V. Nair, B. Partoens, and S. Tarucha,
\textit{Coherent level mixing in dot energy spectra measured by magnetoresonant tunneling spectroscopy of vertical quantum dot molecules},
Phys. Rev. B 81, 245310 (2010).


\bibitem{LMPZ96} R. Laflamme, C. Miquel, J. P. Paz and W. H. Zurek,
\textit{Perfect quantum error correction code},
Phys. Rev. Lett. 77, 198 (1996).
arXiv:quant-ph/9602019v1

\bibitem {Gott96} D. Gottesman,
\textit{A Class of Quantum Error-Correcting Codes Saturating the Quantum Hamming Bound},
Phys. Rev. A 54, 1862 (1996).
arXiv:quant-ph/9604038v2

\bibitem {Shor96} P. W. Shor,
\textit{Fault-tolerant quantum computation},
37th Symposium on Foundations of Computing, IEEE Computer Society Press, 1996, pp. 56-65.
arXiv:quant-ph/9605011v2


\bibitem {KL96} E. Knill, R. Laflamme,
\textit{Concatenated Quantum Codes},
arXiv:quant-ph/9608012v1.

\bibitem {AB96} D. Aharonov, M. Ben-Or,
\textit{Fault Tolerant Quantum Computation with Constant Error},
Proceedings of the 29th Annual ACM Symposium on Theory of Computing, 1997, pp. 176-188.
arXiv:quant-ph/9611025v2

\bibitem {Dya06} M. I. Dyakonov,
\textit{Is Fault-Tolerant Quantum Computation Really Possible?},
Future Trends in Microelectronics. Up the Nano Creek, Wiley (2007), pp. 4-18.
arXiv:quant-ph/0610117v1

\end{thebibliography}
\end{document}